\documentstyle[aps,pra,epsf]{revtex}

\newcommand{\bea}{\begin{eqnarray}}
\newcommand{\eea}{\end{eqnarray}}
\newcommand{\bk}{b^{\dagger}}
\newcommand{\ak}{a^{\dagger}}
\newcommand{\ket}[1]{| #1 \rangle}

\newcommand{\half}{\small \frac{1}{2}}

\begin{document}
\draft
\def\overlay#1#2{\setbox0=\hbox{#1}\setbox1=\hbox to \wd0{\hss #2\hss}#1
-2\wd0\copy1}
\twocolumn[\hsize\textwidth\columnwidth\hsize\csname@twocolumnfalse\endcsname

\title{Fluorescence into flat and structured radiation continua: An
  atomic density matrix without a master equation.}
\author{S\o ren Bay$^{a,b}$, P. Lambropoulos$^{b}$ 
and Klaus M\o lmer$^a$}
\address{
  a. Institute of Physics and Astronomy,
  University of Aarhus,  8000 Aarhus C,  Denmark \\
  b. Max-Planck-Institut f\"ur Quantenoptik\\
  Hans-Kopfermann-Str.\ 1, 85748 Garching, Germany
}
\date{\today}
\maketitle
\begin{abstract}
  We investigate an atomic $\Lambda$-system with one transition
  coupled to a laser field and a flat continuum of vacuum modes and
  the other transition coupled to field modes near the edge of a
  photonic band gap.  The system requires simultaneous treatment of
  Markovian and non-Markovian dissipation processes, but the photonic
  band gap-continuum can not be eliminated within a density matrix
  treatment. Instead we propose a formalism based on Monte-Carlo
  wavefunctions, and we present results relevant to the experimental
  characterization of a structured continuum.
\end{abstract}
\tighten
\vskip0pc]

\begin{figure}
  \begin{center}
    \leavevmode
    \epsfxsize4.5cm
    \epsfbox{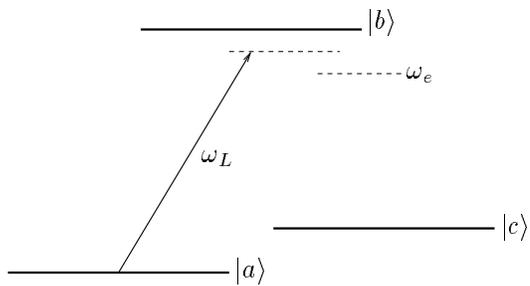}
  \end{center}
\caption{Level scheme}
\label{level.fig}
\end{figure}

With the advent of photonic band gap (PBG)
materials and dispersive media, the mode structure of the 
electromagnetic field can
be tailored in a controllable fashion providing for instance band gaps
or defect modes of various forms \cite{yablo87,johnprl87}.  
The rapidly varying mode structure in the radiation reservoir
invalidates the Born-Markov approximations normally employed 
for a simple quantum system like an atom when this is located inside 
a PBG-material with transition frequency near the edge of the gap.
The reservoir degrees of freedom are thus not
easily eliminated to derive a master equation for the reduced system dynamics.
The main body \cite{johnsd,kurizkisd,baytaioc,baytaipra} of
theoretical works on atomic interactions within PBG materials has
therefore addressed the unitary dynamics in terms of the complete
atom(s)+field wavefunctions.\\  
In this paper, we address a $\Lambda$-system with one
laser-driven transition experiencing a flat vacuum without structure
and the frequency of the other transition near the edge of a PBG.
A unitary wavefunction dynamics
is incompatible with the treatment of the atomic fluorescence
on the ``free''-space transition, and we must seek a way to apply the
simple Markovian properties of this process in the solution of the
complete problem.  
A formulation in terms of Monte-Carlo wavefunctions (MCWF) turns out
to be particularly useful for this purpose. 
This method then also suggests itself as
a means of solving other problems emerging in the overlapping domain
of quantum optics, semiconductors and nano-structures where
dissipation of Markovian and non-Markovian character may co-exist.
Furthermore, our work establishes an application of the MCWF
treatment which goes beyond its conventional correspondence with 
Born-Markov master equations; at no point does such a master equation
appear in this work.  
\\
The $\Lambda$-system is interesting from an experimental point of
view since atoms may be present in their ground state in the
dielectric host, and the dynamics of the interaction with the field
modes in the vicinity of the gap may be studied when the
laser excitation on the ``free''-space transition is turned on. 
In partial analogy with the shelving scheme technique
we note that the fluorescence signal on the
``free''-space transition may serve as to probe details of the
interaction between the atom and the field modes in the PBG
material.
\\ \\
In a PBG one finds a modified
dispersion relation for the photons in the radiation reservoir.
The methods presented in the following apply for any
mode structure and dispersion relation, but to illustrate the method we
employ the isotropic model introduced by John and Wang
\cite{johnprl90,johnprb}. Their dispersion relation for a 
periodic array of dielectric scatterers of radius $a$ and 
index of refaction  $n$, reads (with a separation of $b=2an$
between the scatterers):
\begin{eqnarray}
\omega_k=\frac{c}{4na}\arccos\big[
\frac{4n\cos(2ka(1+n))+(1-n)^2}{(1+n)^2}\big]\nonumber
\label{disprel}
\end{eqnarray}
which leads to a gap centered at the frequency $\omega_0=\frac{\pi
c}{4na}$ . With $n=1.082$ the gap width
$\Delta \omega$ is $0.05\omega_0$ and the upper band edge frequency
$\omega_e$ is given by $\omega_e=\omega_0+\half
\Delta\omega_0=1.025\omega_0$. In the vicinity of the
band edge we have
(effective mass approximation) 
\begin{equation}
\omega_k=\omega_e+A(k-k_0)^2,
\label{ema}
\end{equation}
where $A=-2ac/\sin(4na\omega_e/c)$ and
$k_0=\pi/\left(2a(n+1)\right)$.
\\ \\
We consider a three-level atom with two lower levels $|a\rangle$
and $|c\rangle$ coupled by the electric dipole coupling to 
a common excited level $|b\rangle$, see fig. \ref{level.fig}. 
On the $\ket{a} \leftrightarrow \ket{b}$
transition we apply a laser field, and the atom may decay by
spontaneous emission due to the coupling to a flat 
radiation reservoir. The transition $\ket{b}\rightarrow \ket{c}$ 
is accompanied by
the emission of a photon with frequency in the vicinity of the
photonic band gap edge, and this atomic transition is significantly 
modified by the presence of the dielectric host.

\begin{figure}
 \begin{center}
    \leavevmode
    \epsfysize5cm
    \epsfxsize8.5cm
    \epsfbox{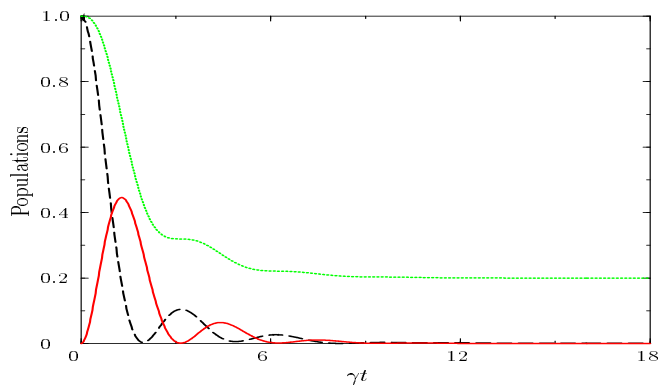}
  \end{center}
\caption{The populations $\pi^0_a(t)$ (dashed
  line), $\pi^0_b(t)$ (solid line) and the norm $P(t)$ (dotted line) 
are plotted as functions of
  time. The parameters chosen
  are: $C^{2/3}=\gamma/3, V_{ab}=\gamma, \omega_e=\omega_b$. }
\label{nojump.fig}
\end{figure}

\begin{figure}
  \begin{center}
    \leavevmode
    \epsfysize5cm
    \epsfxsize8.5cm
    \epsfbox{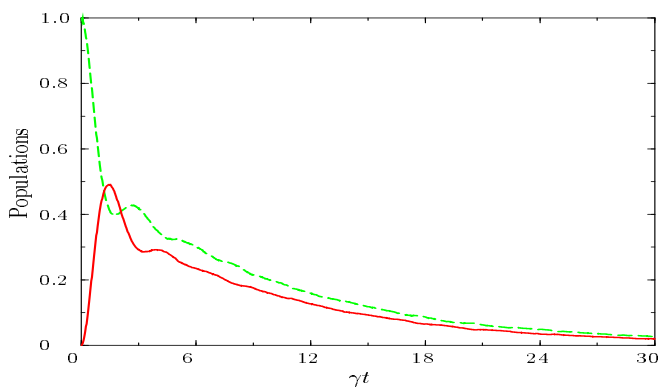}
  \end{center}
\caption{The populations  $\overline{\pi}_{a}(t)$ (dashed
  line) and $\overline{\pi}_{b}(t)$ (solid line) 
  are plotted as functions of time.
  The parameters chosen are the same as in fig. 2. The curves are
  averaged over $10^4$ trajectories.}
\label{mean.fig}
\end{figure}

\begin{figure}
  \begin{center}
    \epsfysize5cm
    \epsfxsize8.5cm
    \epsfbox{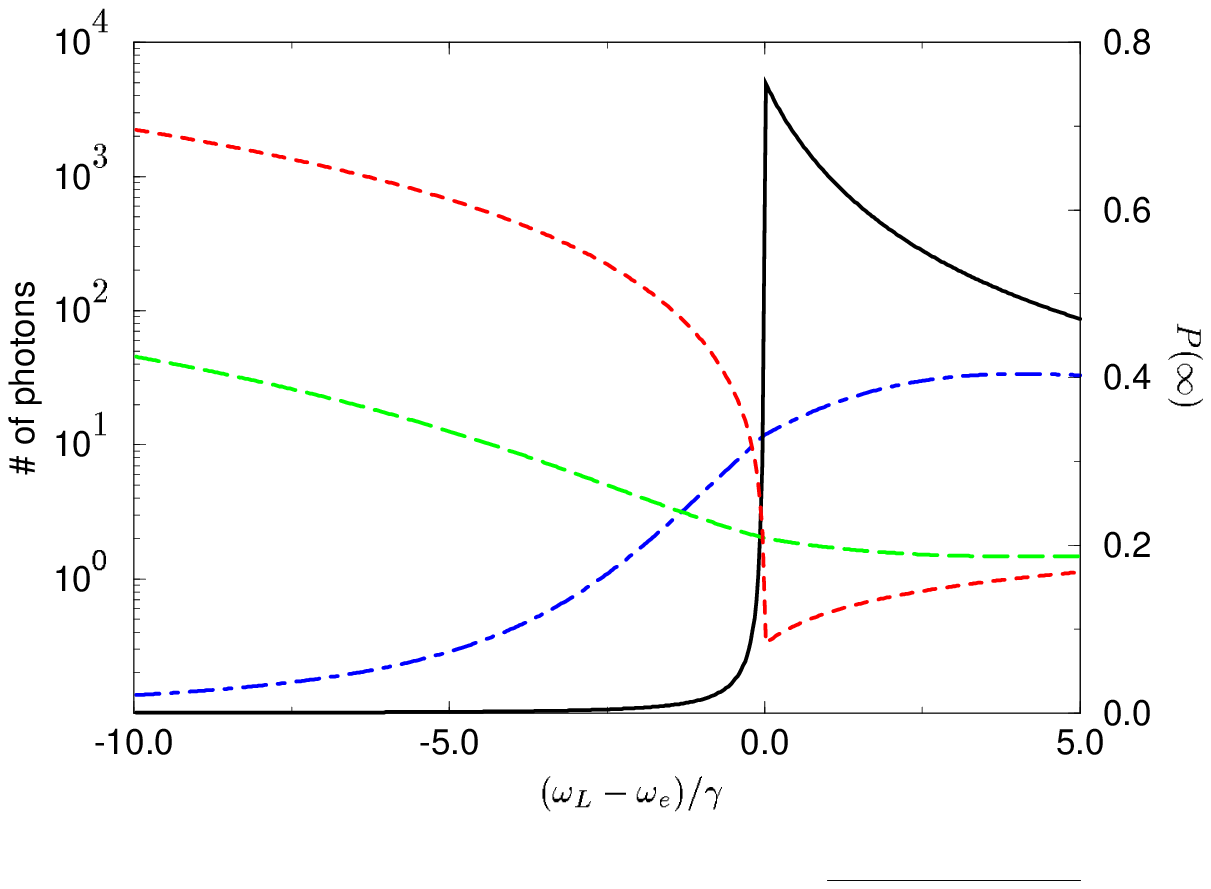}
  \end{center}
\caption{Solid line is $P(\infty)$ for $V_{ab}=\gamma/2 $ and dashed line the
  number of fluorescence photons. The dot-dashed line is 
  $P(\infty)$ for $V_{ab}=3\gamma$ and the long-dashed line the number of
  fluorescence photons. The parameters chosen: $C^{2/3}=\gamma$, 
  $\omega_b=\omega_e$ }
\label{fluor.fig}
\end{figure}

Neglecting the zero-point energies of the field modes, and setting
the atomic energy levels to the values $0,\ \hbar \omega_{b}$ and
$\hbar \omega_{c}$ respectively, we write the
Hamiltonian for the system  $(\hbar=1)$
\begin{eqnarray}
H=\omega_{b}\sigma_{bb}+\omega_{c}\sigma_{cc}
+\sum_{\lambda}\omega_{\lambda}\ak_{\lambda} a_{\lambda}+
\sum_{\lambda}\omega_{\lambda}\bk_{\lambda} b_{\lambda} + V
\label{hamilton}
\end{eqnarray}
where the interaction term in the rotating wave approximation is given by
\begin{eqnarray}
V=i\sum_{\lambda}g_{\lambda}(\ak_{\lambda}\sigma_{ab}-a_{\lambda}\sigma_{ba})
+i\sum_{\lambda}g_{\lambda}(\bk_{\lambda}\sigma_{cb}-b_{\lambda}\sigma_{bc})
\nonumber\\
+ig_{L}(\sigma_{ba}e^{-i\omega_Lt}-\sigma_{ab}e^{i\omega_Lt})
\label{interaction}
\end{eqnarray}
where $\sigma_{ij}$ denote atomic 
dyadic operators $|i\rangle \langle j|$ with $i,j\in\{a,b,c\}$; 
$a_{\lambda},b_{\lambda}$ are the field annihilation operators of the
flat vacuum and PBG vacuum, respectively, and 
the laser field is represented by a semiclassical $c$-number field.
We assume that the coupling to the flat continuum may be treated
by perturbation theory in the usual way, {\it i.e.} an energy shift
(Lamb shift) and a decay rate $\gamma$ may be attributed to the
excited state $|b\rangle$. The Lamb shift is assumed incorporated
in the atomic energy $\omega_{b}$ in eq. (\ref{hamilton}), and the decay rate
describes an incoherent transition mechanism by which atoms in the
excited state $|b\rangle$ decay to the ground state $|a\rangle$.
We shall incorporate the decay mechanism by an
effective non-hermitian Hamiltonian $H_{\mbox{eff}}$. First, we identify
the wavefunction evolution governed by this Hamiltonian, and next, by
appealing to the Monte Carlo wavefunction formalism we shall obtain the
exact evolution of the atomic system.
\\ \\
We apply the resolvent operator defined as \cite{ct92}
\begin{eqnarray}
G(z)=\frac{1}{z-H_{\mbox{eff}}}, \nonumber
\end{eqnarray}
where $z$ is a complex 
Laplace variable and
$H_{\mbox{eff}}$ is the effective non-hermitian 
Hamiltonian of the system given by eqs. (\ref{hamilton}) and
(\ref{interaction}),
but with the sums over $a_{\lambda}$ operators suppressed, and the
replacement $\omega_b \rightarrow \omega_b-i\gamma/2$.\\
With the system initially in state $a$, the resolvent operator
equations read
\begin{eqnarray}
(z-0)G_{aa}(z)&=&1+V_{ab}G_{ba}(z+\omega_L) \nonumber\\
(z-\omega_{\lambda})G_{c_\lambda a}(z)&=&V_{c_\lambda b}G_{ba}(z)
\label{gca}\\
(z-\omega_b+\frac{i\gamma}{2})G_{ba}(z)&=&V_{ba}G_{aa}(z-\omega_L)
+\sum_{\lambda}V_{bc_\lambda}G_{c_\lambda a}(z)\nonumber
\end{eqnarray}
where $V_{ab}=g_L$ and the amplitudes $G_{c_\lambda a}(z)$ pertain to 
the PBG-continuum states $\ket{c}\otimes\ket{1_\lambda}$.
Using eq. (\ref{ema}) in the summation over continuum modes and
turning the summation into an integral, we  get
\begin{eqnarray}
(z-\omega_b+i\gamma/2)G_{ba}(z)=V_{ba}G_{aa}(z-\omega_L)
-\frac{iC G_{ba}(z)}{\sqrt{z-\omega_e}} \nonumber,
\end{eqnarray}
where  the effective dipole coupling to the mode structure is given by
$C=d^2k_0^2\omega_e/(4\pi \varepsilon_0 \sqrt{A})$ \cite{baytaipra},
with $d$ the atomic dipole moment on the $b\leftrightarrow c$ transition.\\
Solving these coupled, algebraic equations for $G_{aa}$ and $G_{ba}$,
we find
\begin{eqnarray}
G_{aa}(z)&=&\frac{(z-\omega_{b}+i\half\gamma)+iC/\sqrt{z-\omega_e}}
{(z-\omega_L)[(z-\omega_{b}+i\half\gamma)+iC/\sqrt{z-\omega_e}]
-|V_{ab}|^2}\label{gaa}\\
G_{ba}(z)
&=&\frac{V_{ba}}{(z-\omega_L)
[(z-\omega_{b}+i\half\gamma)+iC/\sqrt{z-\omega_e}]
-|V_{ba}|^2}\label{gba}
\end{eqnarray}
The dynamics of the system is obtained by inverting the
amplitudes to time domain by means of the inversion integral
for the time evolution operator,
\begin{eqnarray}
{\cal U}(t)=\frac{1}{2\pi i}\int_{\infty+i\epsilon}^{-\infty +i\epsilon}
dz G(z)e^{-izt}\nonumber
\end{eqnarray}
where $\epsilon$ is an infinitesimal small positive quantity.\\
Due to the high order of the polynomial of $z$ in
the denominator and the presence of the square root terms in (\ref{gaa})
and (\ref{gba}), 
it is not easy to apply the residue theorem and to 
obtain the amplitudes in time domain analytically. Instead we
compute the two inversion integrals numerically.
This integration is straightforward, and the calculation yields
for example
the populations of the initial ground state $\ket{a}$ and of the 
excited atomic state $\ket{b}$ as function of time,
$\pi^0_a(t) = |{\cal U}_{aa}(t)|^2,\ \pi^0_b(t) = |{\cal U}_{ba}(t)|^2$.  
We keep track of the norm $P(t)$ of the wavefunction, noting that
it changes only due to the imaginary part of the excited state energy,
and hence 
$\frac{\partial P}{\partial t}|_{loss}=-\gamma\pi^0_b(t)$,
which in integrated form reads
\begin{eqnarray}
P(t)=1-\gamma\int_{0}^tdt'\pi^0_b(t'),
\end{eqnarray}
In fig. \ref{nojump.fig} we show an example of the relevant time dependent 
quantities $P(t),\ \pi^0_a(t)$ and $\pi^0_b(t)$. 
From the figure, it is evident that the populations in the states $\ket{a}$
and $\ket{b}$ approach zero after a transient evolution. There is, however,
a substantial part of the population ($P(\infty) \sim 20\%$) 
which is not lost by
fluorescence on the free-space transition. This population
is transferred to the atomic state $\ket{c}$ associated with the
PBG-continuum, and at any time we have
$\pi^0_c(t)=P(t)-\pi^0_a(t)-\pi^0_b(t)$.
\\ \\
The spontaneous decay on the $\ket{b} \rightarrow \ket{a}$ transition
was treated only as a loss mechanism for the excited state amplitude,
but the atoms are incoherently fed back in the ground state $\ket{a}$,
and from here they are re-excited by the laser. 
In a density matrix formulation, when tracing over the resulting different
photon number states of the flat reservoir, the elimination of the
PBG-modes would be exceedingly difficult, if possible at all.
\\ 
It has been shown that dissipative problems in quantum
mechanics may be solved by stochastic wavefunction equations as an
alternative to master equations
\cite{hjc,moelmerprl,moelmerjosab,dum,hegerfeldtproc}.  In the
``quantum jump'' scheme, one propagates state vectors according to a
non-hermitian Hamiltonian, and at certain instants of time, chosen
according to a random process, this propagation is interrupted by
quantum jump projections of the state vectors (see \cite{moelmerqsopt}
for a recent review).  In the formulations of the method so far, the
continuously propagated state vector is described as the solution of
a Schr\"odinger equation, but, the atomic populations $\pi_i^0(t)$
identified after elimination of the PBG reservoir above may be applied
just as well for the construction of the atomic density matrix.
\\ 
The function $P(t)$ is the norm of the no-jump state vector
\cite{dum} and consequently the probability that a photon has not been
registered in the flat reservoir at time $t$. 
\\ 
The ensemble averaged populations can be found by solution of integral
equations: The population of an atomic state
$\overline{\pi}_i(t)$ is a sum of a term representing the population
given that the atom has not decayed and a term representing the
population given that the latest jump occured at time $t'$,
\begin{eqnarray}
\overline{\pi}_i(t) = \pi^0_i(t) + 
\int_0^t dt' \gamma \overline{\pi}_b(t') \pi_i^0(t-t').
\label{conv}
\end{eqnarray}
Note that $\pi_i^0(t)=\pi_i^0(t)/P(t)\cdot P(t)$ provides the given 
normalized population with the appropiate no-jump weight-factor. 
Eq.(\ref{conv}) must be solved for  $\overline{\pi}_b(t)$ first,
{\it e.g.} by a Laplace transform: $\overline{\pi}_b(z)= \pi_b^0(z)/
(1-\gamma \pi_b^0(z))$, and one may subsequently obtain the other
populations (see also \cite{nienhuis}).\\
The populations can of course also be found by simulations. 
In a single trajectory
one considers the normalized populations $\pi_i(t)=\pi^0_i(t)/P(t)$
(and other density matrix elements if necessary), until a jump occurs
when $P(t)$ equals a random number $\varepsilon$ chosen uniformly on
the interval betwen zero and unity. The quantum jump puts the atom in
the state $\ket{a}$, and from here the evolution starts over again.
In fig. \ref{mean.fig}, we plot
$\overline{\pi}_a(t)$ and $\overline{\pi}_b(t)$ obtained by an average
of $10^4$ stochastic wavefunctions for the same parameters as used in fig.
\ref{nojump.fig}.
\\ \\
Let us comment on the atomic dynamics obtained in fig. \ref{mean.fig}.
After an initial transient evolution, the populations 
$\overline{\pi}_{a,b}(t)$ approach zero in
a non-exponential way. The
fluorescence signal on the $\ket{b}\rightarrow \ket{a}$ transition 
thus vanishes as opposed to the case of a two-level atom in free space. 
In the simulations we note that no jump will occur if the
random number $\varepsilon$ is smaller than $P(\infty)$. 
When this value is non-zero, each
realization only exhibits a limited number of jumps since
eventually the value chosen for $\varepsilon$ will be smaller than
$P(\infty)$. The probability of having exactly $k$ photon emissions
(jumps) in a given simulated trajectory is $(1-P(\infty))^k P(\infty)$,
and the mean number of photons emitted per atom is $P(\infty)^{-1}-1$.
The amplitude ${\cal U}_{c_\lambda a}(t)$ is the inverse Laplace
transform of $G_{c_\lambda a}=V_{c_\lambda
  b}G_{ba}/(z-\omega_\lambda)$.  In the long time-limit only the pole
$z=\omega_\lambda$ contributes and thus ${\cal U}_{c_\lambda
  a}(t)=V_{c_\lambda b}G_{ba}(\omega_\lambda)e^{-i\omega_\lambda t}$
and since $P(\infty)=\pi^0_c(\infty)$ we find
\begin{eqnarray}
P(\infty)=\sum_\lambda |V_{c_\lambda b}G_{ba}(\omega_\lambda)|^2
\nonumber
\end{eqnarray}
The summation can be turned into an integral, which is calculated numerically.
\\
For a $\Lambda$-system in free space
with a branching of the decay from the upper state,
there will also be a finite number of fluorescence photons emitted on the
laser driven transition, corresponding to 
$P(\infty)=\gamma'/(\gamma+\gamma')$ with $\gamma, \gamma'$ being the
decay rates of state $\ket{b}$ 
to the states $\ket{a}$ and $\ket{c}$ respectively, and the total number of
fluorescence photons on the $\ket{b}\rightarrow\ket{a}$ transition is thus
independent of the parameters of the driving field. 
This is different in our case, as seen in fig. \ref{fluor.fig}, where
$P(\infty)$ and the mean number of flourescence photons 
emitted on the free space transition are plotted as function of 
the laser detuning from the PBG edge for different choices of the
laser coupling. For a rather weak laser coupling, the transition to
the atomic state $\ket{c}$ is a Raman-process which is strongly
suppressed when the laser is tuned below the band gap edge since there
are then no resonant PBG-modes for the Stokes photon. A
stronger laser coupling leads to an Autler-Townes splitting of level
$\ket{b}$ and population is then transferred to the PBG-continuum by a
higher order process, removing the step-like character of $P(\infty)$.
The fluorescence signal may hence probe details of the PBG structure.
\\ \\
In conclusion we have demonstrated a technique for the solution of a
problem, for which a Born-Markov master equation does not exist.
We have in a parallel study derived a reduced master
equation for the $\Lambda$-system applying {\it only} the
Born-approximation. The numerical solution of the resulting
non-Markovian master equation yields a very poor agreement with
the exact results presented here thus invalidating the use of the 
Born-approximation. These results will be presented elsewhere.
The specific form of the structured continuum is not essential for our
approach, but it is important that only one photon states of the
PBG-continuum appear (a possible slow decay from state $\ket{c}$ back
to $\ket{a}$ can only be treated if we may assume that the photon in
the PBG continuum escapes before the atom is reexcited to level
$\ket{b}$). The simulations and the analytical expression (\ref{conv})
are simplified by the fact that all jumps put the atom in the same
state. 
Our formalism, however, is perfectly capable of treating more general
systems with branching of the decay from state $\ket{b}$ to multiple
states $\ket{a_j}$. This implies that the index $a$ is replaced by the
set of indices $a_j$ with the corresponding enlargement of the set of
equations (\ref{gca}) and (\ref{conv}). 
The no-jump evolution and the associated
delay function, following a jump to a given level $a_j$, are then readily
computed. \\
The MCWF technique
has been applied to non-Markovian problems through the solution of
Markovian master equations for enlarged model systems \cite{imamoglu}.
Our situation, however, is different, since without ever having a master
equation we have, by recourse to conditioned wavefunction dynamics,
been able to obtain the atomic density matrix. 
We anticipate that by combination of the ideas in
ref. \cite{imamoglu} and in this paper, a wide class of non-Markovian
problems may become tractable.

\bibliography{$HOME/tex/paper/tai/pra/photonic}
\bibliographystyle{prsty}

\end{document}